\documentclass{article}
\usepackage{official/spconf,amsmath,amssymb,graphicx,booktabs,multirow,hyperref,xcolor}
\usepackage{colortbl}
\definecolor{tablegroupbg}{gray}{0.93}
\hypersetup{hidelinks,pdftitle={Beyond Token Revision: Investigating Mask-and-Replace Diffusion for Zero-Shot Text-to-Speech},pdfauthor={Hounsu Kim, Joonyong Park, Yuki Saito, Satoru Fukayama, Juhan Nam}}

\newcommand{\mask}{\textsc{mask}}

\newcommand{\policyresult}[2]{#1~(#2)}

\newcommand{\CurrentMainCommitRows}{%
\rowcolor{tablegroupbg}[0pt][\tabcolsep]
\multicolumn{5}{@{}l}{\textit{CosyVoice3}\quad $K{=}$6,561; 25~Hz} \\
AR-LM & $1.1/.4/1^{\mathrm{off}}$ & 12.01\hphantom{ (5.97)} & 4.1273 & 0.7008 \\
Mo-DLM & $1.2/.6/.8^{P}$ & \underline{6.05} (5.97) & \underline{4.2320} & \underline{0.7140} \\
DeMaR (ours) & $1.3/.2/1^{P}$ & \textbf{4.95} (5.23) & \textbf{4.2760} & \textbf{0.7173} \\
\midrule
\rowcolor{tablegroupbg}[0pt][\tabcolsep]
\multicolumn{5}{@{}l}{\textit{FlexiCodec}\quad $K{=}$32,768; $\approx 8.3$~Hz} \\
AR-LM & $1.3/.4/.8^{\mathrm{off}}$ & 7.25\hphantom{ (5.38)} & 3.8067 & 0.6747 \\
Mo-DLM & $1/0/1^{P}$ & \underline{5.39} (5.38) & \textbf{3.8791} & \textbf{0.6942} \\
DeMaR (ours) & $1.1/0/.8^{P}$ & \textbf{4.89} (5.03) & \textbf{3.8820} & \textbf{0.6938} \\
}

\newcommand{\SeedControlRows}{%
\rowcolor{tablegroupbg}[0pt][\tabcolsep]
\textit{CosyVoice3} & $1.3/0/.95^{P}$ & $1.2/.6/1^{P}$ & $1.2/.6/1^{P}$ & \\
No rep. & \policyresult{7.17}{6.87} & \policyresult{6.88}{7.03} & \policyresult{6.68}{6.92} & 6.91\,$\pm$\,0.24 \\
Rep., no loss & \policyresult{6.28}{5.73} & \policyresult{\textbf{5.35}}{5.29} & \policyresult{5.20}{5.28} & 5.61\,$\pm$\,0.58 \\
Rep. + loss & \policyresult{\textbf{5.08}}{5.07} & \policyresult{5.41}{5.54} & \policyresult{\textbf{4.59}}{4.80} & \textbf{5.02}\,$\pm$\,0.41 \\
\midrule
\rowcolor{tablegroupbg}[0pt][\tabcolsep]
\textit{FlexiCodec} & $1.1/0/1^{P}$ & $1.1/0/.8^{E}$ & $1.1/0/.9^{E}$ & \\
No rep. & \policyresult{5.49}{5.61} & \policyresult{5.68}{5.61} & \policyresult{6.09}{6.02} & 5.75\,$\pm$\,0.31 \\
Rep., no loss & \policyresult{5.26}{5.27} & \policyresult{5.39}{5.53} & \policyresult{5.86}{5.88} & 5.50\,$\pm$\,0.31 \\
Rep. + loss & \policyresult{\textbf{4.94}}{5.03} & \policyresult{\textbf{4.71}}{4.78} & \policyresult{\textbf{5.30}}{5.49} & \textbf{4.98}\,$\pm$\,0.30 \\
}

\title{Beyond Token Revision: Investigating Mask-and-Replace Diffusion\\for Zero-Shot Text-to-Speech}
\name{Hounsu Kim$^1$, Joonyong Park$^{1,2}$, Yuki Saito$^{2,3\dagger}$, Satoru Fukayama$^{3\dagger}$, Juhan Nam$^{1\dagger}$\thanks{\small $^{\dagger}$Co-corresponding authors.}}
\address{$^1$KAIST, Republic of Korea \qquad $^2$The University of Tokyo, Japan\\
$^3$National Institute of Advanced Industrial Science and Technology (AIST), Japan}

\begin{document}
\ninept
\maketitle

\begin{abstract}
Unlike autoregressive models, discrete diffusion–based models for zero-shot text-to-speech generate speech tokens in parallel and can revisit earlier predictions.
Mask-and-replace training extends mask-only training by randomly replacing some tokens, and its gains are commonly attributed to self-correction, the ability to revise previously generated tokens.
However, exposure to randomly perturbed context during training may itself improve generation, raising the question of whether these gains require inference-time token revision.
To investigate this question, we use DeMaR, which combines mask-and-replace training with confidence-ranked mask-only sampling while preserving the total training corruption probability.
Trained from scratch on LibriTTS, DeMaR achieves lower word error rates (WER) than autoregressive and mask-only diffusion baselines using the same speech tokenizer.
This advantage persists when each token remains unchanged after first being unmasked.
Matched training conditions on two heterogeneous speech tokenizers show that both noisy-context augmentation and replacement supervision improve WER under this restriction.
These findings demonstrate training-side benefits of replacement beyond enabling inference-time token revision.

\end{abstract}


\section{Introduction}
\label{sec:intro}
Zero-shot text-to-speech (TTS) generates speech from text using reference speech to specify the speaker's voice, without speaker-specific fine-tuning~\cite{du2024cosyvoice,wang2024maskgct}.
Speech language models (SpeechLMs) perform this task by generating discrete speech tokens that a waveform decoder converts to audio~\cite{wang2023valle}.
Autoregressive language models (AR-LMs) generate tokens sequentially, whereas \emph{discrete diffusion language models} (DLMs) recover clean speech-token sequences from corrupted inputs through repeated parallel predictions~\cite{sahoo2024mdlm}.
With linguistic content supplied by the input text, this non-autoregressive generation offers a way to reduce reliance on preceding token predictions and potentially limit error accumulation~\cite{liu2026texttotalk}.

In the forward process, mask-only training masks some target tokens, whereas mask-and-replace training also replaces some with randomly sampled alternatives~\cite{gu2022vqdiffusion,vonrutte2025gidd}.
Both approaches train the model to reconstruct the original sequence.
Reconstruction loss at replaced positions (\emph{replacement supervision}) trains the model to recover the original tokens from randomly substituted ones, potentially supporting self-correction at inference.
Beyond serving as reconstruction targets, these replaced positions also provide noisy context for predicting masked tokens.
We hypothesize that this \emph{noisy-context augmentation} can improve generation even without replacement supervision or inference-time revision.

To test this hypothesis, we use \emph{DeMaR} (\emph{decoupled mask-and-replace}), which combines mask-and-replace training with confidence-ranked mask-only sampling for conditional speech-token generation.
Following D3RM~\cite{kim2025d3rm}, DeMaR uses replacement only during training. At inference, masks alone carry the combined training probability of masking and replacement.
This mask-only sampling procedure allows a \emph{commit-only} variant in which each token remains unchanged once it is first unmasked.
Keeping the trained model and masking schedule fixed, we use this variant to test whether the gains persist without inference-time revision.

\begin{figure*}[t]
\centering
\includegraphics[width=\textwidth]{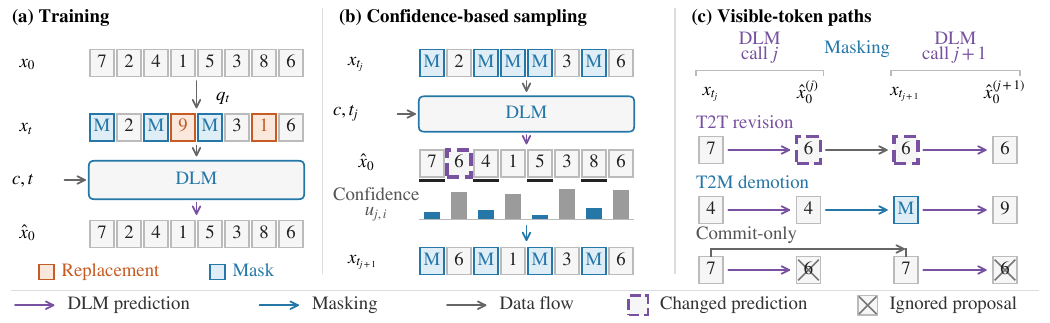}
\vspace{-12pt}
\caption{\textbf{DeMaR training and sampling.}
Panels (a) and (b) show the proposed method; (c) the three paths that distinguish the update policies we compare.
(a) Mask-and-replace training.
(b) Confidence-ranked mask-only sampling; the $k_j$ lowest-confidence positions are underlined and masked.
(c) Visible-token paths: revisiting permits token-to-token (T2T) revision and token-to-mask (T2M) demotion, whereas commit-only preserves visible tokens and fills only masked positions.
Token values and confidence scores are illustrative; $j$ indexes denoiser calls.}
\label{fig:method}
\end{figure*}

Our contributions are summarized as follows:
\begin{list}{$\bullet$}{\setlength{\leftmargin}{1em}\setlength{\topsep}{0pt}\setlength{\itemsep}{0pt}\setlength{\parsep}{0pt}}
\item \textbf{Benefits beyond inference-time revision.}
DeMaR retains its WER advantage over mask-only DLMs (Mo-DLMs) under commit-only sampling.
We also distinguish the effects of direct token revision from those of confidence-based remasking.
\item \textbf{Training-side contributions.}
Matched training conditions show that noisy-context augmentation improves WER without replacement supervision or inference-time revision, while replacement supervision provides further gains under commit-only sampling.
\end{list}
Our comparison covers two heterogeneous tokenizers: CosyVoice3's fixed-rate tokens~\cite{du2025cosyvoice3} and FlexiCodec's variable-rate content and span sequences~\cite{li2025flexicodec}.

\section{Related Work}
\label{sec:background}
\textbf{Inference-time token revision.}
Standard masked decoders keep a token once it is unmasked: MDLM copies visible tokens unchanged~\cite{sahoo2024mdlm}, and confidence-ranked generators such as MaskGCT, MaskGIT, and SoundStorm never remask them~\cite{wang2024maskgct,chang2022maskgit,borsos2023soundstorm}.
Token-to-mask (T2M) remasking returns a visible token to \mask{}; ReMDM and \emph{Mask, Sample, Revise} apply it without retraining~\cite{wang2025remdm,ferreira2026masksamplerevise}, while PALLE uses it in a separate refinement stage~\cite{yang2025palle}.
Token-to-token (T2T) overwriting changes a visible token directly; GIDD and SCOPE pair this operation with training on incorrect visible tokens~\cite{vonrutte2025gidd,yu2026scope}.
\emph{Remask, Don't Replace} compares the two and finds T2M more effective, mainly for correction-trained models~\cite{yao2026remask}.

\textbf{Mask-and-replace training.}
Mask-and-replace training exposes the denoiser to incorrect visible tokens, both as reconstruction targets and as context for predicting masked tokens.
VQ-Diffusion and Diffsound adopt it partly to avoid accumulating errors, and InstructTTS applies it to TTS~\cite{gu2022vqdiffusion,yang2023diffsound,yang2024instructtts}.
GIDD applies it to text and suggests that self-correction can occur during denoising~\cite{vonrutte2025gidd}, CDLM explicitly supervises corrupted visible tokens to support targeted refinement~\cite{zhang2026cdlm}, and SCOPE instead trains on self-generated predictions matched to its sampler~\cite{yu2026scope}.

\textbf{Decoupling training and sampling corruption.}
D3RM trains with masks and random replacements but samples with masks only for piano transcription~\cite{kim2025d3rm}.
It assigns the combined corruption probability to masks.
This improves accuracy over mask-and-replace sampling, and under this sampling, mask-and-replace training still outperforms mask-only training.
DeMaR adapts this training--sampling separation to speech tokens, using confidence-ranked masking rather than D3RM's posterior sampler.
We test whether DeMaR's gains persist without inference-time revision and whether noisy-context augmentation and replacement supervision each contribute.

\section{DeMaR and Matched Comparisons}
\label{sec:method}
We formulate zero-shot TTS as text-conditioned speech-token continuation, following the prefix-conditioning setups of MaskGCT and VALL-E~\cite{wang2024maskgct,wang2023valle}.
The conditioning sequence $c$ consists of the concatenated prompt and target text tokens, followed by the uncorrupted prompt speech tokens.
Let $\mathbf{x}_0=(x_{0,1},\ldots,x_{0,L})\in\mathcal V^L$ be a clean target sequence of $L$ speech tokens from vocabulary $\mathcal V$.
Given a corrupted sequence $\mathbf{x}_t$, conditioning $c$, and diffusion time $t\in[0,1]$, the denoiser with parameters $\theta$ predicts a categorical distribution at each target position, assigning probability $p_\theta(x_{0,i}\mid\mathbf{x}_t,c,t)$ to the ground-truth token.

\subsection{Training corruption and objective}
\label{sec:corruption}
At a shared diffusion time $t$, each target token $x_0$ is independently corrupted according to
\begin{equation}
q(x_t\mid x_0)=a_t\delta_{x_0}+m_t\delta_{\mask}+r_t\mathcal U(\mathcal V),
\label{eq:corruption}
\end{equation}
where $\delta_x$ is a point mass, $\mask\notin\mathcal V$ is the mask symbol, and $\mathcal U(\mathcal V)$ samples each vocabulary symbol with probability $1/|\mathcal V|$, including the original token.
Here $a_t$, $m_t$, and $r_t$ are the probabilities of retaining, masking, and randomly replacing a token, respectively.
Following GIDD~\cite{vonrutte2025gidd}, we set $h(t)=\tfrac12\sqrt{t(1-t)}$ and use
\begin{equation}
(a_t,m_t,r_t)=\frac{(1-t,t,h(t))}{1+h(t)}.
\label{eq:schedule}
\end{equation}
These probabilities sum to one, with clean inputs at $t=0$ and fully masked inputs at $t=1$.
The matched Mo-DLM baseline masks tokens with probability $\rho_t=m_t+r_t$ and retains them otherwise.
The two models therefore select positions for corruption with the same probability but apply different corruption operations.

For each training example, we uniformly sample one of $T$ diffusion timesteps evenly spaced in $[0,1]$ and corrupt the target input at that time.
DeMaR and Mo-DLM reconstruct the original tokens using the same weighted cross-entropy over target positions $\mathcal I=\{1,\ldots,L\}$:
\begin{equation}
\mathcal L=-\frac{\sum_{i\in\mathcal I}w_i\log p_\theta(x_{0,i}\mid\mathbf{x}_t,c,t)}
{\sum_{i\in\mathcal I}w_i}.
\label{eq:loss}
\end{equation}
The loss weight $w_i$ is 1 at positions selected for masking or replacement and 0.1 otherwise.

For CosyVoice3, this formulation applies to a single sequence of fixed-rate speech tokens.
FlexiCodec instead uses paired sequences of content and span tokens, with each span indicating the number of consecutive pre-merge tokens combined into the corresponding content token.
The two streams are corrupted independently at the same sampled diffusion time.
We apply Eq.~\ref{eq:loss} to each stream and sum the normalized losses.

\subsection{Confidence-based sampling}
\label{sec:sampling}
DeMaR and Mo-DLM share a confidence-based token selection and masking procedure.
For Mo-DLM, the sampling mask budget follows its mask-only training probability $\rho_t$.
DeMaR instead trains with both masking and replacement but allocates their combined probability $\rho_t=m_t+r_t$ entirely to masking during sampling.
We call this change of corruption process at inference \emph{decoupled sampling (DS)}.

Generation starts from a fully masked target sequence of length $L$.
Let $N$ be the number of denoiser calls, i.e., function evaluations (NFE), indexed by $j=0,\ldots,N-1$.
At call $j$, we use training timestep $s_j$ and its corresponding diffusion time $t_j$:
\begin{equation}
t_j=\frac{s_j-1}{T-1},\qquad
s_j=\left\lceil\frac{T(N-j)}{N}\right\rceil.
\label{eq:sampling_time}
\end{equation}

At denoiser call $j$, we select a token proposal $\hat{x}_{0,i}$ at position $i$ from the sampling distribution $\pi_{j,i}\in[0,1]^{|\mathcal V|}$, where $\sum_{v\in\mathcal V}\pi_{j,i}(v)=1$.
We assign confidence $u_{j,i}\in\mathbb R$ using either the proposal's probability under $\pi_{j,i}$ or the negative entropy of $\pi_{j,i}$~\cite{li2026omnidiffusion}.
For both models, the number of masked positions in the next input follows the corruption probability $\rho_{t_{j+1}}$:
\begin{equation}
k_j=\operatorname{round}\!\left[L\rho_{t_{j+1}}\right],
\qquad j=0,\ldots,N-2.
\label{eq:sampling_count}
\end{equation}
For the final call, we set $k_{N-1}=0$.
Under the default revisiting policy, the denoiser proposes tokens at every target position.
We select the $k_j$ lowest-confidence positions as $R_j\subseteq\mathcal I$.
For $j<N-1$, the next input is constructed from the predicted clean sequence by
\begin{equation}
x_{t_{j+1},i}=\begin{cases}
\mask,&i\in R_j,\\
\hat{x}_{0,i},&i\notin R_j.
\end{cases}
\label{eq:ds_update}
\end{equation}
This revisiting policy permits two operations (Fig.~\ref{fig:method}(c)).
A \emph{token-to-token (T2T) revision} changes a previously visible token to a different vocabulary token that remains unmasked in the next state.
\emph{Token-to-mask (T2M) demotion}~\cite{yao2026remask} remasks a visible position selected by the low-confidence ranking.
When that position is later unmasked, its token may be the same as or different from the token before masking.
Under revisiting, the schedule determines how many positions are masked, while confidence rankings determine which positions are masked at each call.

\subsection{Controlling inference-time revision}
\label{sec:controls}
To test whether DeMaR's advantage over Mo-DLM depends on inference-time revision, we replace revisiting with \emph{commit-only} sampling (Fig.~\ref{fig:method}(c)).
Both policies run the denoiser on the full sequence, but commit-only uses new proposals and confidence scores only at currently masked target positions, $M_j=\{i\in\mathcal I:x_{t_j,i}=\mask\}$.
At call $j$, it fills the $|M_j|-k_j$ highest-confidence positions in $M_j$ with their proposed tokens and leaves the remaining $k_j$ positions masked.
Commit-only therefore disables both T2T revision and T2M demotion.

To test whether either revision pathway improves generation beyond commit-only sampling, we evaluate them separately.
\emph{T2T-only} allows visible tokens to change directly but selects masks only within $M_j$.
\emph{T2M-only} selects masks across all target positions but preserves the current values of visible tokens not selected for masking.
All policies use the same mask count $k_j$.

\subsection{Separating augmentation from replacement supervision}
\label{sec:training_controls}
To separate noisy-context augmentation from direct replacement supervision, we compare three training conditions.
Let $\mathcal M\subseteq\mathcal I$ and $\mathcal R\subseteq\mathcal I\setminus\mathcal M$ denote the masking and replacement-candidate positions sampled under Eq.~\ref{eq:corruption}.
The remaining positions are $\mathcal O=\mathcal I\setminus(\mathcal M\cup\mathcal R)$.
In \emph{no replacement}, tokens at $\mathcal R$ remain unchanged and receive zero loss weight.
\emph{Replacement, no loss} replaces tokens at $\mathcal R$ with uniformly sampled tokens but assigns no loss to these positions.
\emph{Replacement + loss} (DeMaR) uses the same corrupted inputs and adds unit-weight reconstruction loss at $\mathcal R$.

All three conditions use loss weights 1 on $\mathcal M$ and 0.1 on $\mathcal O$, replacing the denominator in Eq.~\ref{eq:loss} with
\begin{equation}
Z=|\mathcal M|+|\mathcal R|+0.1|\mathcal O|.
\label{eq:control_normalization}
\end{equation}
Including $|\mathcal R|$ keeps the denominator unchanged when replacement supervision is added.
Unlike the no-replacement condition, Mo-DLM masks both $\mathcal M$ and $\mathcal R$ and supervises them with unit weight.

All three conditions use the same sampling mask budget, determined by $\rho_t=m_t+r_t$, as in Table~\ref{tab:main}.
Comparing them under revisiting and commit-only sampling tests whether the benefits of augmentation and replacement supervision require inference-time revision.

\section{Experimental Setup}
\label{sec:setup}
\subsection{Data and model training}
We train speech-token generators from scratch on 585 hours of LibriTTS~\cite{zen2019libritts} using the frozen tokenizers and waveform decoders of CosyVoice3~\cite{du2025cosyvoice3} and FlexiCodec~\cite{li2025flexicodec}.
All models share a backbone with 12 layers, a hidden dimension of 768, and 12 attention heads, totaling 84.95M parameters excluding vocabulary-dependent embeddings.
They share text preprocessing and tokenization and are trained with AdamW at a learning rate of $4\times10^{-4}$ on two NVIDIA RTX 4090 GPUs per run.

AR-LMs use causal attention and teacher forcing to predict speech tokens and EOS, whereas DLMs use bidirectional attention to reconstruct corrupted tokens; both condition on the full transcript.
During DLM training, we corrupt tokens after a clean prefix covering 10--40\% of the sequence, or the entire sequence with probability 0.15.
We select DLM checkpoints by dev-clean WER using oracle target-token counts.
Each condition in Section~\ref{sec:training_controls} is trained with three seeds; the other tables report seed 1.

\subsection{Decoding hyperparameter tuning}
\label{sec:decoding}
For the system comparison, we tune each model's decoding hyperparameters by dev-clean WER using Gaussian-process Bayesian optimization with expected improvement (GP-EI)~\cite{snoek2012bayesian}.
We search on 1,000 dev-clean utterances and rerank the top four candidates on 3,000 separate dev-clean utterances.
We search temperatures 0, 0.2, 0.4, 0.6; top-$p$ 0.8, 0.9, 0.95, 1.0; and repetition penalties~\cite{keskar2019ctrl} from 1.0 to 1.6 in steps of 0.1.
For AR-LMs, we also consider repetition-aware sampling (RAS), either disabled or using (window, threshold) pairs $(5,0.2)$, $(20,0.1)$, or $(20,0.2)$.
We stop after three consecutive GP checks in which every unevaluated candidate has an expected WER reduction below 0.02 percentage points and less than a 10\% probability of improving the best observed WER by at least 0.1 percentage points.

DLMs use 32 denoiser calls with constant temperature; we compare the two confidence rules in Section~\ref{sec:sampling}.
Repetition penalty, temperature, and top-$p$ filtering modify $\pi_{j,i}$, the distribution used for both token proposals and confidence scores.
In Table~\ref{tab:main}, commit-only uses the same checkpoint and decoding settings as revisiting.
For each tokenizer and seed in Table~\ref{tab:supervision}, all three training conditions use the no-replacement model's selected decoding settings.

\subsection{Evaluation protocol and metrics}
We evaluate all 4,835 LibriTTS test-clean utterances.
Within each tokenizer, all systems share the evaluation pairs, text conditioning, full reference speech-token prefix, and frozen waveform decoder.
AR-LMs terminate at EOS, whereas DLMs use target-token counts predicted by OmniVoice's prompt-based length rule~\cite{omnivoice2026duration}.
We measure intelligibility using utterance-averaged WER between HuBERT-Large CTC~\cite{hsu2021hubert} transcripts of generated speech and normalized written references.
UTMOS22-strong~\cite{saeki2022utmos} estimates naturalness, while WavLM-Large~\cite{chen2022wavlm} measures speaker similarity to the reference speech (SIM).

\section{Results and Analysis}
\label{sec:results}

\begin{table}[t]
\centering\small
\setlength{\tabcolsep}{3pt}
\caption{Generation quality with RP/temperature/top-$p$ settings ($\tau=0$: greedy).
Parentheses give commit-only WER.
Superscripts denote DLM confidence (P: probability; E: entropy+RP) or the AR RAS window (numerals; threshold 0.2, off disables RAS).}
\label{tab:main}
\label{tab:commit}
\begin{tabular*}{\columnwidth}{@{}l@{\hspace{3pt}}c@{\extracolsep{\fill}}r@{\extracolsep{0pt}\hspace{5pt}}c@{\hspace{5pt}}c@{}}
\toprule
System & RP/$\tau$/$p$ & \multicolumn{1}{@{}c@{\extracolsep{0pt}\hspace{5pt}}}{WER (\%) $\downarrow$} & UTMOS $\uparrow$ & SIM $\uparrow$ \\
\midrule
\CurrentMainCommitRows
\bottomrule
\end{tabular*}
\end{table}

DeMaR outperforms both AR-LM and Mo-DLM in WER (Table~\ref{tab:main}), reducing Mo-DLM’s WER from 6.05 to 4.95\% for CosyVoice3 and from 5.39 to 4.89\% for FlexiCodec.
Both DLMs also outperform AR-LM in UTMOS and SIM.
For CosyVoice3, DeMaR raises UTMOS from 4.127 to 4.276 and SIM from 0.701 to 0.717 relative to AR-LM.
Differences between the two DLMs are more pronounced in WER than in UTMOS or SIM; for FlexiCodec, both achieve approximately 3.88 UTMOS and 0.694 SIM.
We next examine whether DeMaR's WER gains require inference-time revision and how noisy-context augmentation and replacement supervision contribute.

\subsection{Contribution of inference-time revision}
\label{sec:revision_results}
Switching from revisiting to commit-only raises DeMaR's WER by only 0.29 percentage points for CosyVoice3 and 0.14 for FlexiCodec (Table~\ref{tab:commit}).
Under commit-only sampling, DeMaR still outperforms Mo-DLM: 5.23 versus 5.97\% for CosyVoice3 and 5.03 versus 5.38\% for FlexiCodec.
Mo-DLM's WER decreases slightly under the same switch, from 6.05 to 5.97\% and from 5.39 to 5.38\%, respectively.
Thus, DeMaR's advantage does not require inference-time revision, motivating the following analysis of its training components.

\subsection{Noisy-context augmentation and replacement supervision}
\label{sec:replacement_benefits}
Table~\ref{tab:supervision} focuses on WER, as UTMOS and SIM vary by at most 0.08 and 0.006, respectively, within each tokenizer and seed.
Unlike Mo-DLM, the no-replacement model leaves replacement-candidate positions unchanged and unsupervised during training (Section~\ref{sec:training_controls}).
Under commit-only sampling, adding random replacements without supervising them reduces seed-1 WER from 6.87 to 5.73\% for CosyVoice3 and from 5.61 to 5.27\% for FlexiCodec.

Adding replacement supervision further lowers seed-1 commit-only WER to 5.07\% for CosyVoice3 and 5.03\% for FlexiCodec.
The two replacement-trained models share the same training-input distribution, so this comparison measures the effect of supervising replaced positions.
This comparison does not separate the benefit of reconstructing replaced tokens from that of increasing the number of supervised positions.
Augmentation lowers WER in all three seeds for both tokenizers under both sampling policies.
With revisiting, mean WER decreases from $6.91\pm0.24$ to $5.61\pm0.58$\% for CosyVoice3 and from $5.75\pm0.31$ to $5.50\pm0.31$\% for FlexiCodec.
Replacement supervision further reduces the means to $5.02\pm0.41$\% and $4.98\pm0.30$\%, respectively, improving WER in two of three CosyVoice3 seeds and all three FlexiCodec seeds under both sampling policies.
The exception is CosyVoice3 seed 2, where WER increases by 0.06 percentage points with revisiting and 0.24 with commit-only.
Thus, augmentation consistently helps across the tested seeds, while replacement supervision provides additional but less uniform gains.

{\emergencystretch=2em
We next examine whether learning to reconstruct replaced tokens translates into additional gains from T2T revision or T2M demotion at inference (Table~\ref{tab:revision_paths}, CosyVoice3).
Relative to commit-only, T2T-only raises WER from 6.87 to 7.20\% without replacement and from 5.72 to 6.27\% with replacement but no loss.
With replacement supervision, this penalty disappears: T2T-only yields 5.02\% WER versus 5.07\% for commit-only.
T2M-only differs from commit-only by at most 0.04 percentage points for all three models.
With replacement supervision, WER spans only 5.02--5.08\% across all four policies, with neither revision pathway substantially improving on commit-only sampling.
Together with the commit-only improvements in Table~\ref{tab:supervision}, this suggests that learning to reconstruct replaced tokens benefits generation beyond performing token revision at inference.
\par}

\begin{table}[t]
\centering\small
\setlength{\tabcolsep}{3pt}
\caption{Matched training conditions, one column per training seed, each using that seed's own no-replacement selection.
Rep. denotes replacement; Rep.\ + loss is DeMaR.
Shaded rows give each seed's setting in Table~\ref{tab:main} notation.
Cells give revisiting WER (\%) $\downarrow$; parentheses give commit-only WER.}
\label{tab:supervision}
\begin{tabular*}{\columnwidth}{@{\extracolsep{\fill}}lrrrr@{}}
\toprule
Condition & seed 1 & seed 2 & seed 3 & mean\,$\pm$\,sd \\
\midrule
\SeedControlRows
\bottomrule
\end{tabular*}
\end{table}

\begin{table}[t]
\centering\small
\setlength{\tabcolsep}{3pt}
\caption{WER (\%) with T2T revision and T2M demotion controlled separately, using seed 1's CosyVoice3 setting in Table~\ref{tab:supervision}.
Settings follow Table~\ref{tab:main} notation.}
\label{tab:revision_paths}
\begin{tabular*}{\columnwidth}{@{\extracolsep{\fill}}lrrrr@{}}
\toprule
Condition & Commit-only & T2T-only & T2M-only & Revisiting \\
\midrule
\rowcolor{tablegroupbg}[0pt][\tabcolsep]
\multicolumn{5}{@{}l}{\emph{CosyVoice3: revision pathways}\quad $1.3/0/.95^{P}$} \\
No rep. & 6.87 & 7.20 & 6.87 & 7.17 \\
Rep., no loss & 5.72 & 6.27 & 5.76 & 6.27 \\
Rep.\ + loss & 5.07 & 5.02 & 5.08 & 5.08 \\
\bottomrule
\end{tabular*}
\end{table}

\section{Conclusion}
We tested whether the gains from mask-and-replace training depend on inference-time token revision.
When each token remains unchanged after first being unmasked, DeMaR retains its WER advantage over mask-only diffusion, with WER increasing by only 0.29 percentage points for CosyVoice3 and 0.14 for FlexiCodec.
Noisy-context augmentation improves WER without replacement supervision in all three seeds for both tokenizers, under both revisiting and commit-only sampling.
Replacement supervision provides further gains in two of three CosyVoice3 seeds and all three FlexiCodec seeds under both policies.
These findings show that noisy-context augmentation and replacement supervision can improve generation even when generated tokens are never revised.

\clearpage
\section{Acknowledgements and Use of Generative AI Disclosure}
This work was supported by JST BOOST JPMJBY24C9, the AIST policy-based budget project “R\&D on Generative AI Foundation Models for the Physical Domain.”, and the National Research Foundation of Korea (NRF) grant funded by the Korea government (MSIT) (No. RS-2023-00222383).

\noindent \textbf{Use of Generative AI Disclosure:} OpenAI's ChatGPT and Codex assisted manuscript drafting and editing.

\bibliographystyle{official/IEEEbib}
\begingroup
\small
\bibliography{references}
\endgroup
\end{document}